\documentclass[aps,prc,preprint,groupedaddress]{revtex4-2}
\usepackage{graphicx}
\usepackage{dcolumn}

\usepackage{bm}

\begin{document}

\title{Direct reaction contribution in the very low energy $^{19}$F(p,$\alpha_0$) reaction}
\author{Lalit Kumar Sahoo}
\author{Chinmay Basu}
\email[]{chinmay.basu@saha.ac.in}

\affiliation{Saha Institute of Nuclear Physics,HBNI, 1/AF Bidhannagar, Salt lake, Kolkata - 700064, INDIA}
\date{\today}
\begin{abstract}
The direct reaction component of the $^{19}$F(p,$\alpha_{0}$) reaction at E$_{cm}$=180-600 keV is studied for the data that has become very recently available. This component has been found to be significant in the present work  using the direct pickup model in framework of the DWBA formalism and indicate the strong cluster structure of $^{19}$F.
\end{abstract}


\maketitle

\section{Introduction}

The$^{19}$F(p,$\alpha_0$) reaction at E$_{cm}$=27-94 keV is one of the two destruction paths (the other being $^{19}$F(p,$\gamma$)) of fluorine, synthesised in AGB stars \cite{il}. The reaction rate is very important to know as stellar models fail to describe the overabundance of fluorine in these sites. The reaction populating the ground state of $^{16}$O is found to be dominant at astrophysical energies and detection of the emergent alpha particle, eliminating the large background is required. There are several direct measurements \cite{mor86,clpl57,cuzzo80,iso59,cura74,bruer59}at energies much above the Gamow window and in this scenario, theoretical extrapolation to the Gamow region is the only solution as for many other thermonuclear reactions. However, there is an interesting aspect about this reaction in terms of it's mechanism, that poses an uncertainty in the extracted reaction rates. It is not very well understood from previous studies, to what extent the measured cross-sections can be analysed in terms of only R-matrix theory. This assumes that the reaction proceeds by the formation of compound nuclear resonances and all the direct measurements were analysed in terms of the R-matrix phenomenology. 
A work of Herndl et al \cite{h} has contradicted this assumption, who analysed the angular distributions of this reaction in the framework of the direct reaction model. Herndl analyzed the angular distribution data of Lorenz-Wirzba et al.\cite{lw}, measured at 250, 350 and 450 KeV, in the framework of the zero range DWBA theory for direct pickup and calculated the astrophysical S-factor. 
Immediately after the results of Herndl et al.\cite{h}, Y.Yamashita et al.\cite{y} analysed the same data of Wirzba et al. in the frame work of finite range DWBA. The results of Y.Yamashita was significantly different from Herndl. It showed that an exchange process in addition to the direct pick up picture is necessary to explain the experimental angular distributions. The difference in the two conclusions drawn by Herndl and Yamashita was attributed to the zero and finite range DWBA calculations \cite{l}. 
However this finding was not included in the NACRE compilation \cite{ang99} due to systematic errors in the data of Wirzba et al \cite{lw}.
and the direct reaction mechanism remain unattended in the extraction of the astrophysical S-factor so far.
Very recently, the cross-sections for the $^{19}$F(p,$\alpha_{0}$ reaction has been measured by Lombardo et al \cite{l} at E$_{cm}$=180-600 keV. This is the lowest energy measurement of the $^{19}$F(p,$\alpha_0$) cross-section and R-matrix analysis has been done and S-factor extracted without considering the direct reaction component. 
The present work, revisits the problem raised by Herndl et al and investigates the role of direct reaction in the $^{19}$F(p,$\alpha$) reaction using the recent and more precise Lombardo data \citep{l}.  

\section{The reaction mechanism of the $^{19}$F(p,$\alpha_0$) reaction} 
$^{19}$F(p,$\alpha_0$) is a positive Q-value reaction ( Q= + 8.114 MeV) and is energetically possible to occur at very low projectile energies. The electromagnetic decay channel therefore competes with the alpha emission channel. The gamma emission occurs mainly from the resonant states of $^{20}$Ne and excitation function can be fitted using the R-matrix phenomenology. However, due to the cluster structure of $^{19}$F, it is likely that the direct reaction process become significant even at low energies and cannot be ignored in the extraction of the S-factor. Two possible picture for the direct
reaction may be possible.In one, $^{19}$F has t+$^{16}$O cluster components and the proton picks up the triton and forms the alpha particle and $^{16}$O recoils i.e 
\begin{equation}
p+^{19}F(t + ^{16}O)\rightarrow \alpha(p+t)+^{16}O
\end{equation}
 Alternatively, $^{19}$F can be modeled as $\alpha + ^{15}N$  and the alpha in the target is exchanged with the proton of the projectile.
\begin{equation}
p+^{19}F(\alpha+^{15}N)\rightarrow^{16}O(p+^{15}N)+\alpha
\end{equation} 
Besides, the alpha particle can be emitted from the excited $^{20}$Ne compound nucleus, in the form of statistical evaporation i.e
\begin{equation}
p+^{19}F\rightarrow^{20}Ne^{*}\rightarrow ^{16}O+\alpha   
\end{equation}

\section{Results and Discussions}
Since both the works of Herndl et al. and Y.Yamashita et al. provide conflicting results in the frame work of direct reaction model, it is necessary to reanalyse the new angular distribution data of \cite{l} in terms of direct reaction formalism. In the present work, we perform first the pickup  process calculations using the code FRESCO \cite{sf} version 3.1 in the framework of FR-DWBA and ZR-DWBA formalism. 
In table 1, we give the potential parameters used in the present work, the potentials used by Herndl et al. and Yamashita et al. are given for comparison. The results of the FRESCO \cite{sf} calculation are compared to the Wirzba data in fig 1 and Lombardo data in fig 2.  The Wirzba data was extracted using software GetData Graph Digitizer \cite{G} as it was unavailable in the EXFOR database. The 4s and 1s state is used for the bound state $t+ ^{16}$O and $p+t$ respectively. The spectroscopic amplitude used for best fitting data lies between 0.5-0.7 in FR-DWBA.In case of ZR-DWBA calculation, the code need only spectroscopic amplitude for target bound state and independent of projectile bound state spectroscopic amplitude. So the only best fitting spectroscopic amplitude for the target bound state  $t+ ^{16}$O lies between 0.1-0.2.The Zero range coupling constant taken in the calculation is \cite{M.W} $$D_0^2~= 4.53\times10^5~~ MeV^2~fm^3$$.

As evident from figure 1 and 2, the FR-DWBA calculations  show a more satisfactory explanation of the experimental data as compared to ZR-DWBA particularly at lower energies.The Mehta et al.\cite{M.K} potential is used for the exit channel $\alpha+^{16}$O. However the results improve with the change of the depth of exit channel potential and the energy dependence relation of potential is given in the table of our present work.The angular distribution and the DWBA calculations for Wirzba et al. data\cite{lw} and Lombardo et al. data\cite{l} at respective energies are shown in figure ~\ref{wirzba et al.} and ~\ref{Lombardo et al.} respectively.The FR-DWBA calculation shows good agreement with experimental data up to 500 KeV.However the 616 KeV data is well explained by the total cross section which is combination of direct and exchange process.The Woods-Saxon form of bound state parameters for exchange process $^{15}N+p$ and $^{15}N+\alpha$ is taken from Y.Yamashita et al.\cite{y}.The bound state parameter $r_0$ value of $^{15}N+p$ is adjusted to 1.4 fm to give an exchange angular distribution and which is ultimately added to direct part for the case of 616 KeV. The present calculations are therefore satisfactory except at 616 keV where the resonance reaction is also important.
\begin{table}[hp]
  \centering
  \caption{Optical potential parameters}
  \small
  \begin{ruledtabular}
    \begin{tabular}{p{0.15\textwidth}p{0.27\textwidth}p{0.27\textwidth}p{0.27\textwidth}} 
    
      \textbf{System} & Herndl et. al & Yamashita et. al & This Work\\
      \colrule
      p+$^{19}$F &&& \\
      Real part & V$_o$=56.5 - 2E$_{lab}$ & V$_o$=56.5 - 2E$_{lab}$ & V$_o$=60.0  \\
       & r$_v$=1.12 , a$_v$=0.48 & r$_v$=1.12 , a$_v$=0.48& r$_v$=1.20 , a$_v$=0.5\\
       \\
       Imaginary &W$_o$=0.625+1.5E$_{lab}$ &W$_o$=0.625+1.5E$_{lab}$&W$_o$=0.625+1.5E$_{lab}$\\
      part & r$_w$=1.25 , a$_w$ = 0.55 & r$_w$=1.25 , a$_w$ = 0.55 & r$_w$=1.55 , a$_w$ = 0.55\\ 
       &rc=1.20& rc=1.20 &rc=1.20\\
       
       \colrule
       $\alpha$+$^{16}$O \\
        Real part& Double folding & V$_o$=150.0& V$_o$=175 - 92E$_{lab}$\\
       &  potential&r$_v$=0.723 , a$_v$=0.48 & r$_v$=0.723 , a$_v$=0.48 \\
       \\
       Imaginary & W$_o$=2 & W$_d$ =3.0& W$_d$ =3.0\\
       part & r$_w$=1.25 , a$_w$ = 0.6&r$_d$=0.723 , a$_d$=0.5&r$_d$=0.723 , a$_d$=0.5\\
       & r$_c$=1.25 &r$_c$=1.40 \cite{M.K} &r$_c$=1.40  \\
       
       \colrule
       $^{19}$F=$^{16}$O+t & &V$_o$=108.96 & V$_o$=108.96\\
       &&r$_v$=1.31 , a$_v$=0.65 &r$_v$=1.31 , a$_v$=0.65\\
       &&r$_c$=1.31 & r$_c$=1.31  \cite{h} \\ 
       \colrule
      $\alpha$=p+t & &V$_o$=60.99 & V$_o$=60.99\\
      & &r$_v$=1.43 , a$_v$=0.3 &r$_v$=1.43 , a$_v$=0.3\\
       &&r$_c$=1.43 &r$_c$=1.43 \cite{S}\\
       
    \end{tabular}
    \end{ruledtabular}
    \footnote*{ V and W are in MeV ,r and a in fm }\\ 
    
\end{table}

\begin{figure}[h]
\begin{center}
\begin{tabular}{cc}
\includegraphics[width=0.35\textwidth]{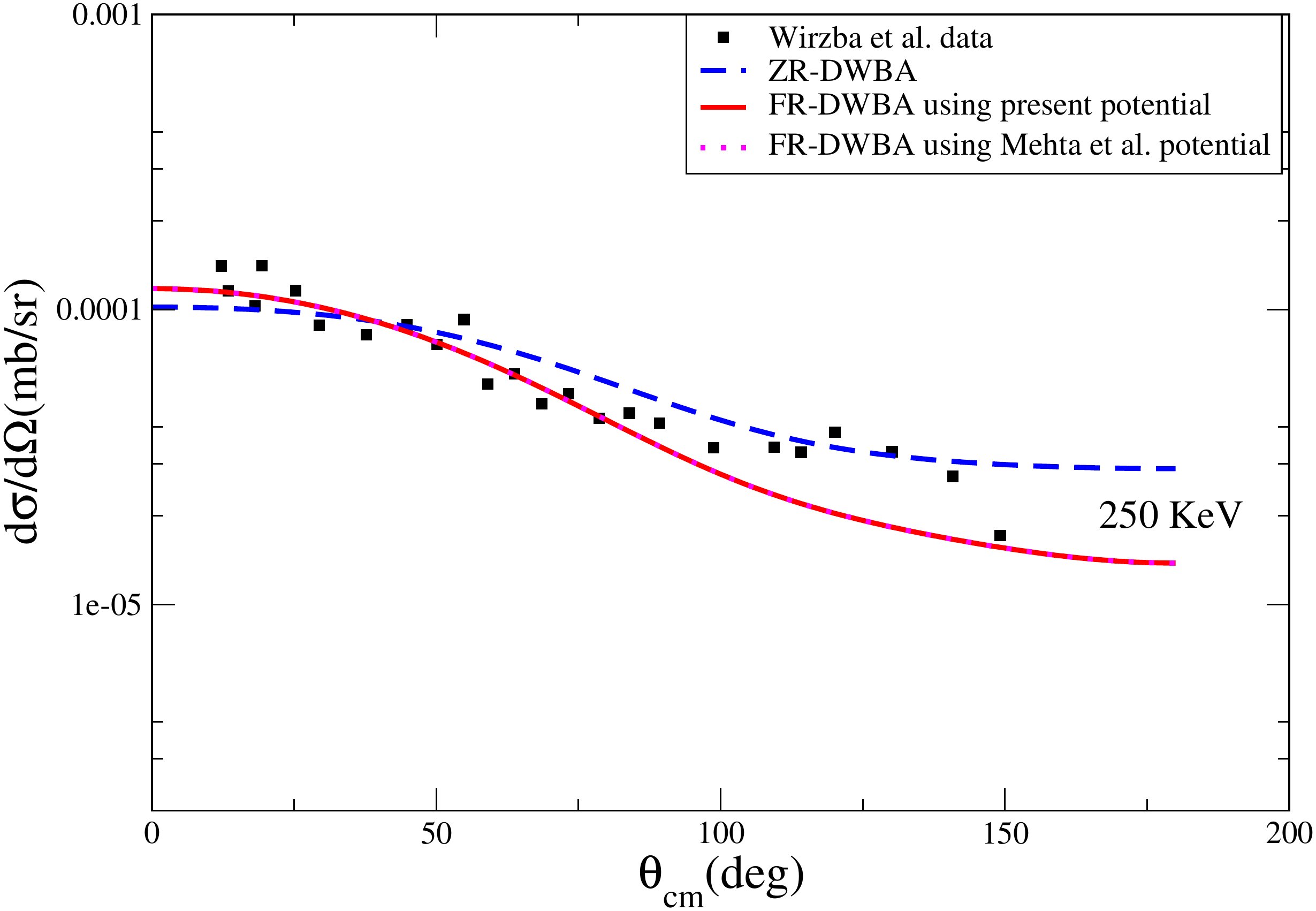}
\includegraphics[width=0.35\textwidth]{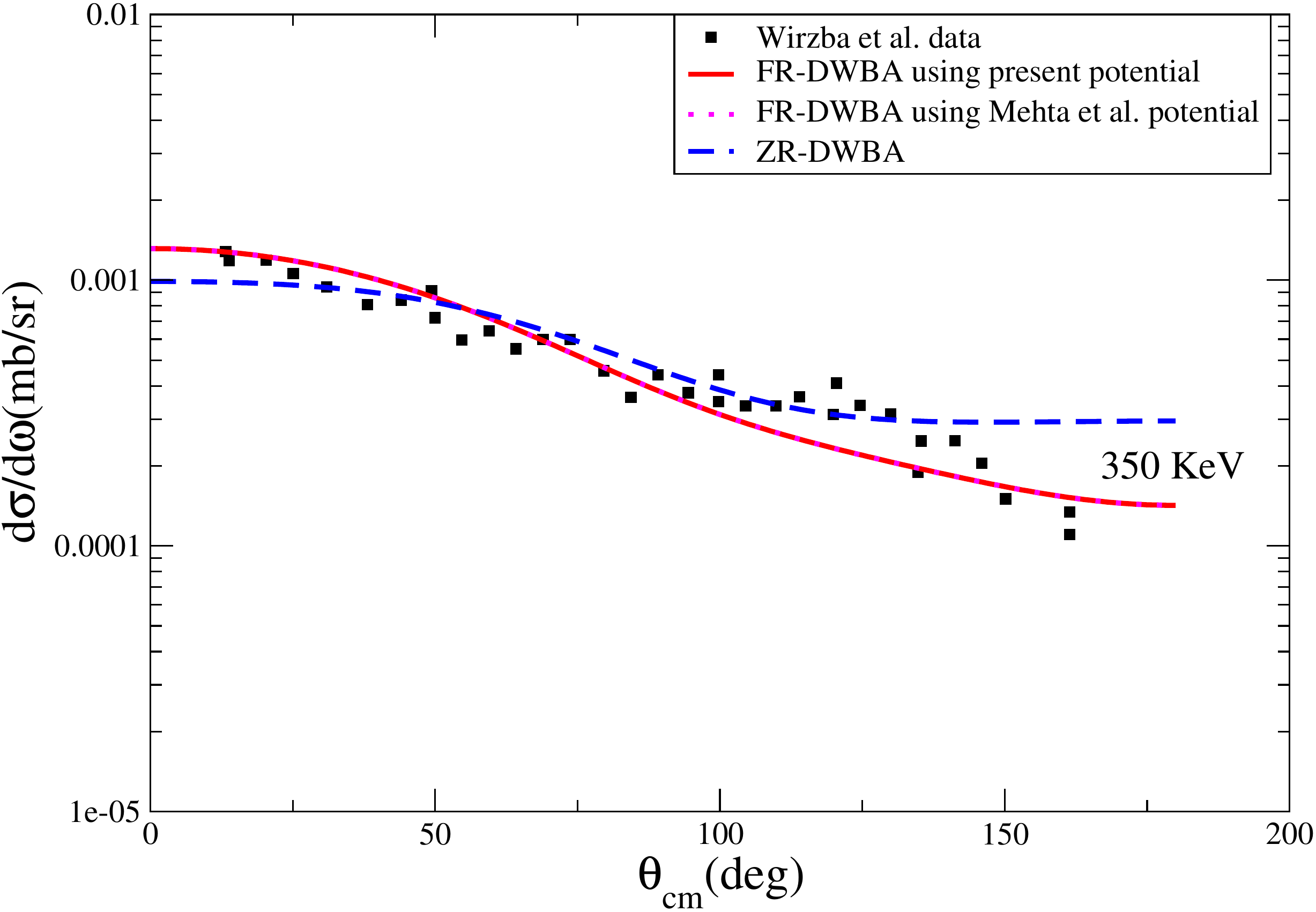}\\
\includegraphics[width=0.35\textwidth]{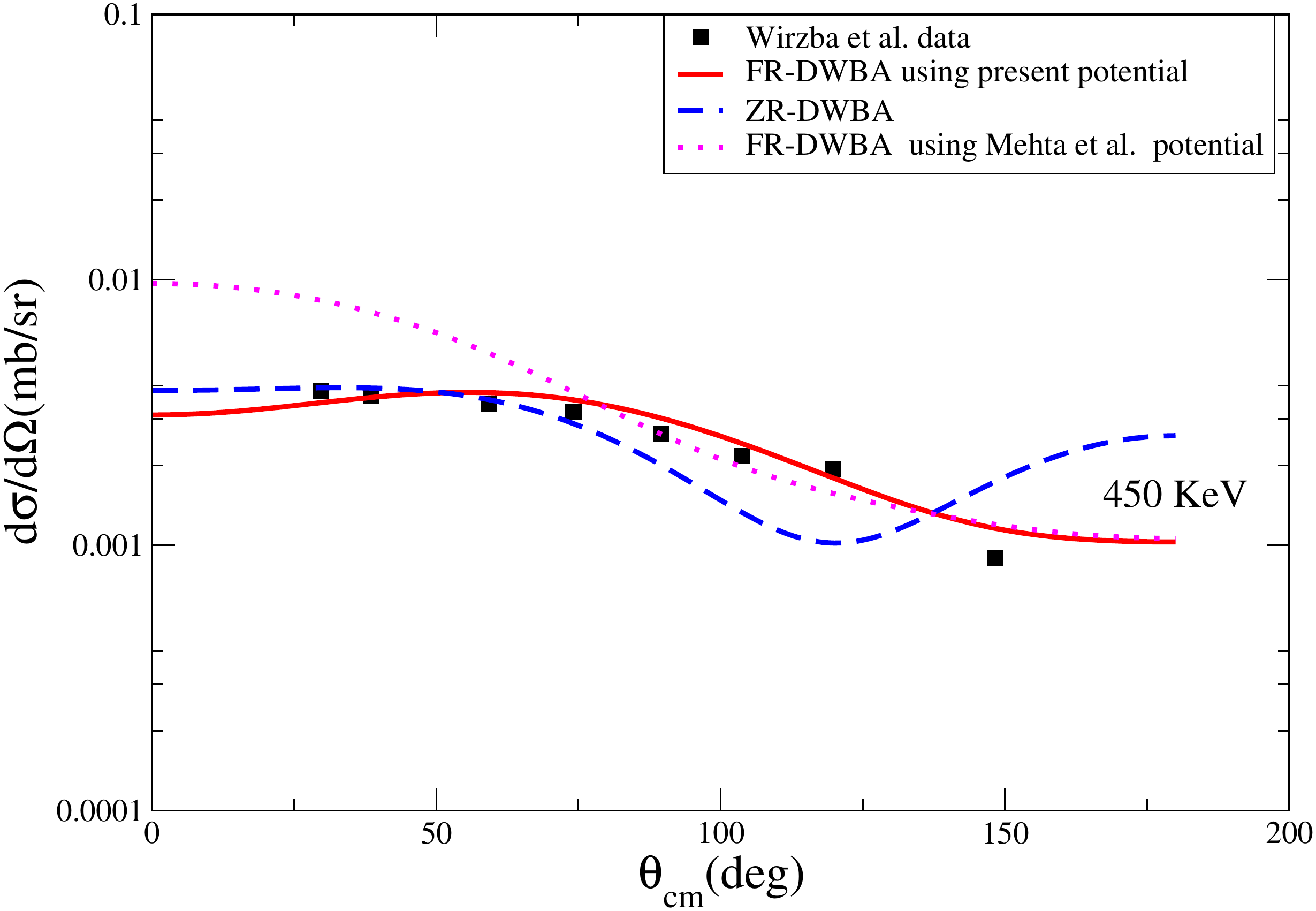}
\end{tabular}
\caption{Angular distribution for the reaction $^{19}$F(p,$\alpha_0)^{16}O$ at 250, 350 and 450 KeV.The  experimental data are taken from \cite{lw}}.
\label{wirzba et al.}
\end{center}
\end{figure}
\begin{figure}[h!]
\begin{center}
\begin{tabular}{cc}
\includegraphics[width=0.35\textwidth]{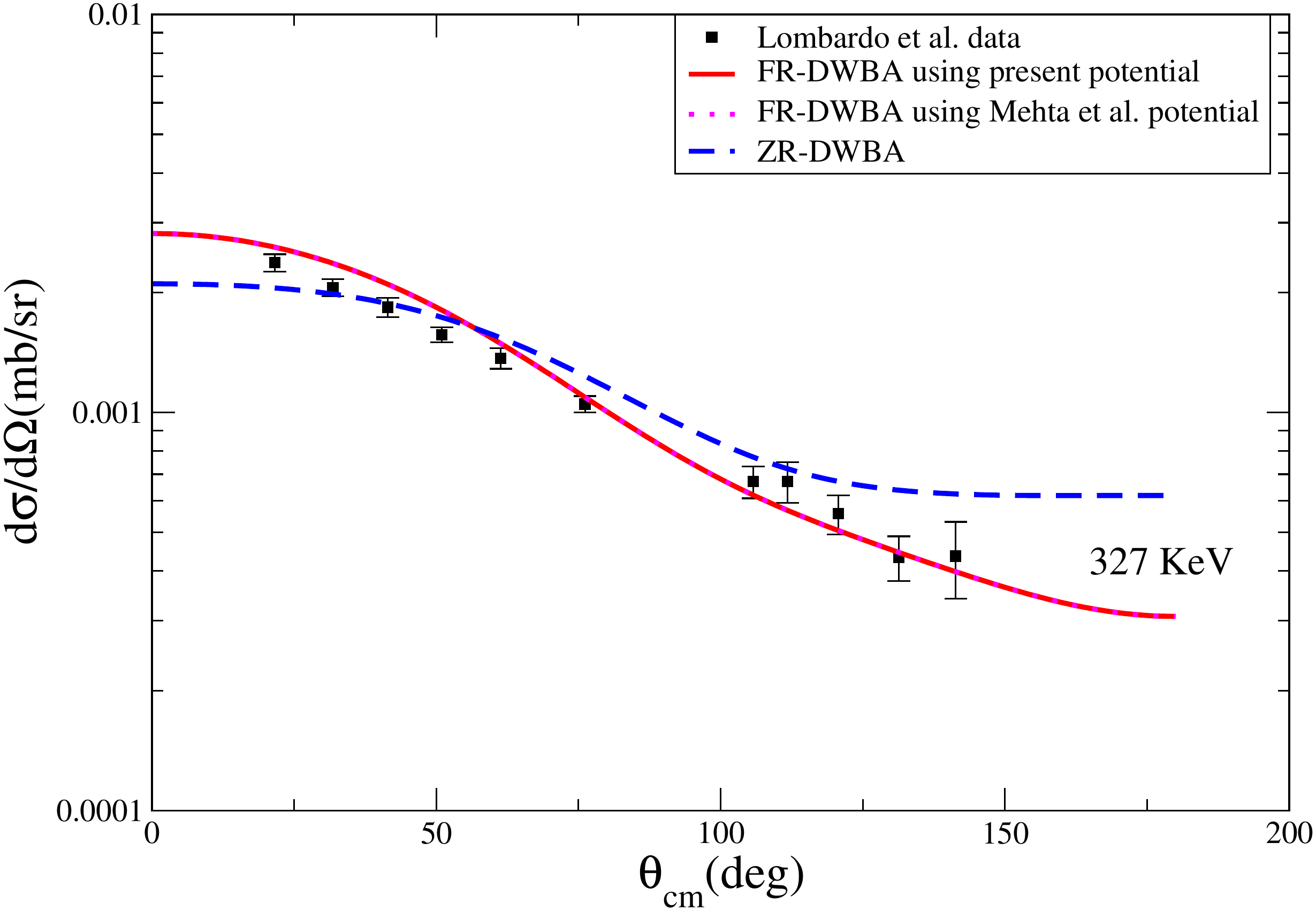}
\includegraphics[width=0.35\textwidth]{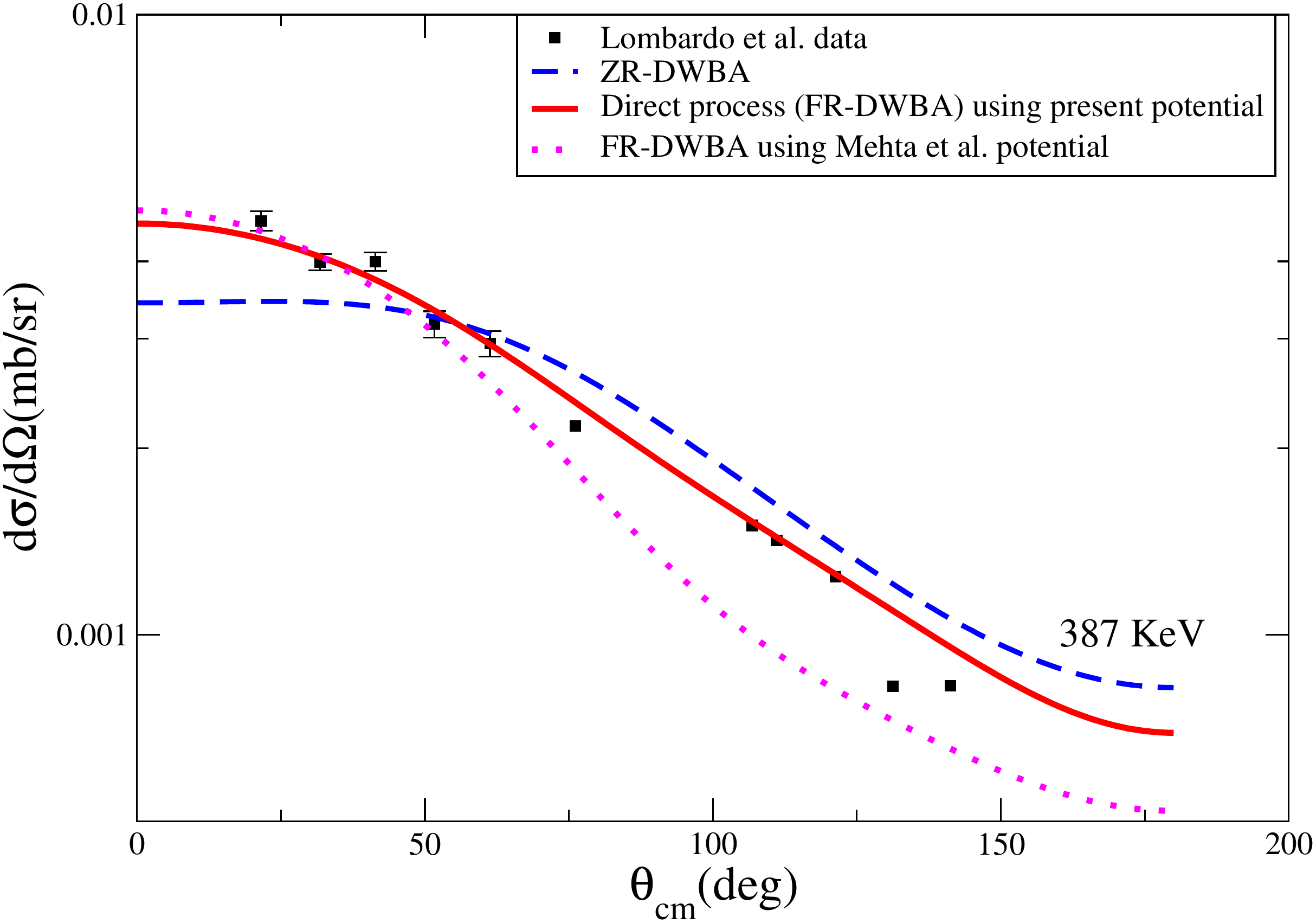}\\
\includegraphics[width=0.35\textwidth]{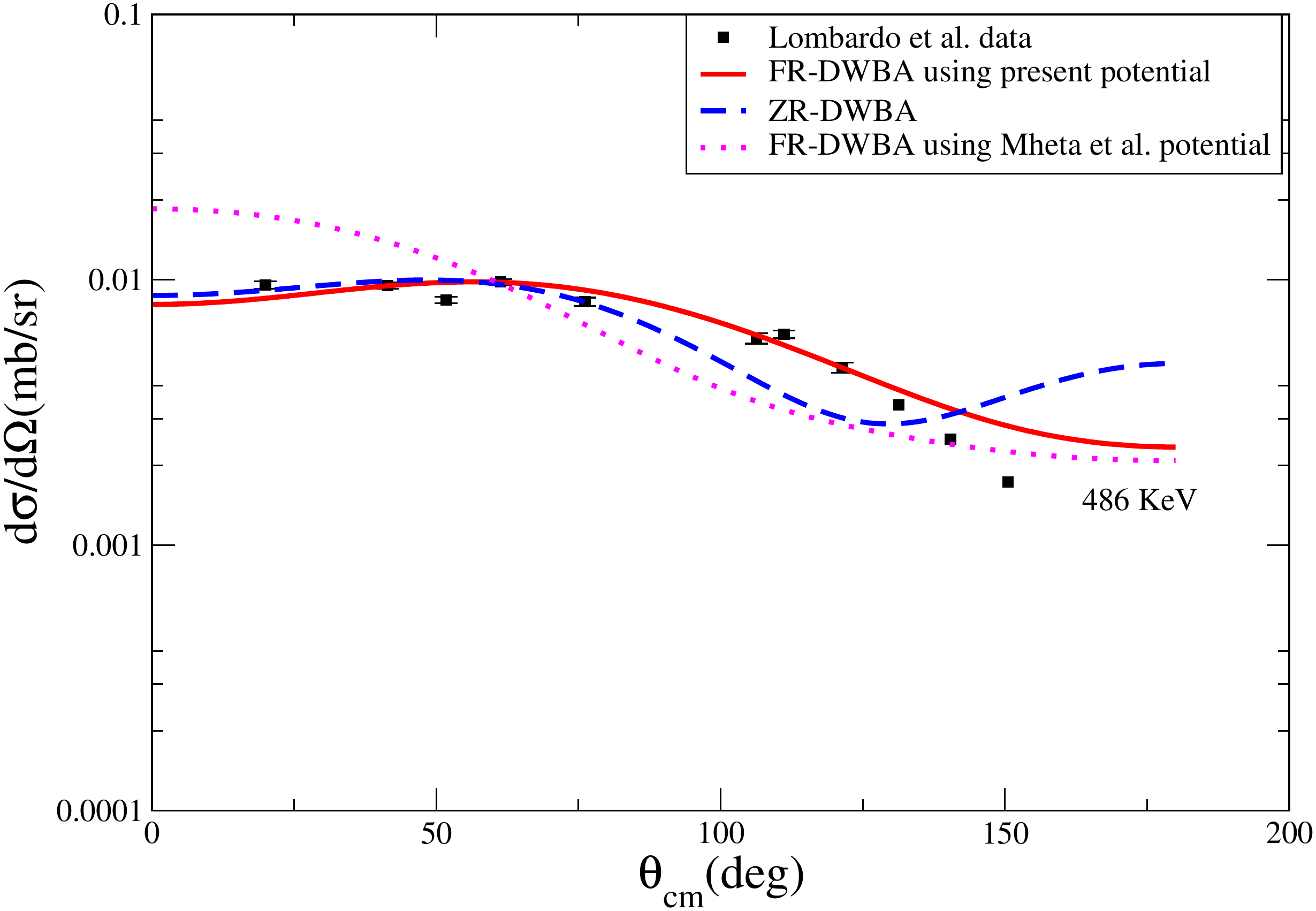}
\includegraphics[width=0.35\textwidth]{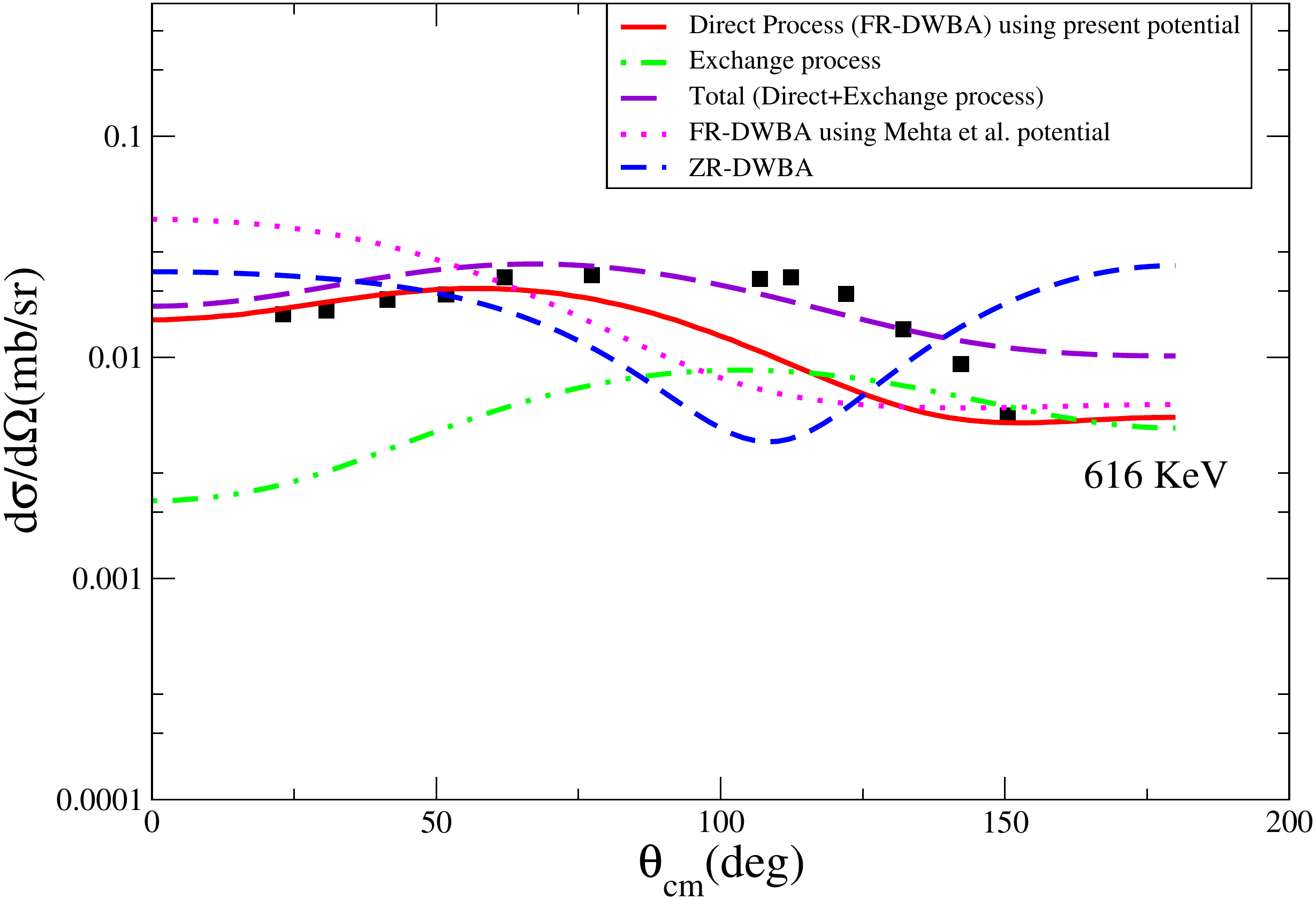}\\
\end{tabular}
\caption{Angular distribution for the reaction $^{19}$F(p,$\alpha_0)^{16}O$ at 327,387,486 and 616 KeV.The  experimental data are taken from \cite{l}}
\label{Lombardo et al.}
\end{center}
\end{figure}
\ \\
\ \\
\ \\
\ \\
\ \\
\section{R-matrix phenomenology}
In the previous section the results of the direct reaction model calculations are discussed with respect to the most recent angular distribution data of the $^{19}$F(p,$\alpha_0$) reaction at low energies. The results of the direct reaction model are strongly dependent on the interaction potentials and the type of transfer mechanism such as "direct" or "exchange" type. On the other hand the $^{19}$F(p,$\alpha_0$) excitation functions have been thoroughly analysed by Lombardo et al.\cite{l} in terms of R-matrix theory to explain the resonances. However,in those analysis the angular distribution have not been fitted by R-Matrix theory.They have used Legendre polynomial fit to explain the angular distribution.It is interesting to calculate the angular distribution in terms of R-matrix theory, particularly when direct reaction results are conflicting.The result of R-matrix fitting and angular distribution is carried out in Azure 2 \cite{Rmatrix} considering only four background state.The inclusion or exclusion of $^{20}Ne$ state does not make any difference to the result.The ANC of $^{16}O$ ground state is fixed at 37460.25 $fm^{-0.5}$.The proton and $\alpha-$ channel of four background state $0^+$,$1^-$,$2^+$ and $3^-$ are freely varied to get the best fit.The large ~MeV order width of the state signifies to smaller lifetime which ultimately shows significant contribution of direct part below coulomb barrier.The fitted parameters are shown in TABLE-II for all the energy.

\begin{table}[hp]
\begin{ruledtabular}
\caption{Fitted parameters for  Lombardo et al.\cite{l} data and Wirzba data \cite{lw}}
\centering

 \begin{tabular}{p{2cm} p{2cm} p{3cm} p{3cm} p{2cm} } 
 
 E (KeV)& $E_x$ (MeV) & $J^{\pi}$   & $\Gamma_\alpha$ (MeV) & $ \chi^2/N$ \\ 
  \colrule
  & 20 & $0^+$ &  1253.758 &   \\ 
 
 327 & 20& $1^-$  & 1209.840 &  0.11 \\
 & 20 & $2^+$ & 2.666 & \\
 & 20 & $3^-$ & 7.465 & \\
 \colrule
  & 20 & $0^+$ &  1298.841 &   \\ 
 
 387 & 20& $1^-$  & 990.143 & 0.92   \\
 & 20 & $2^+$ & 9.529 & \\
 & 20 & $3^-$ & 0.323 & \\
 \colrule
  & 20 & $0^+$ &  1258.866 &   \\ 
 
 486 & 20& $1^-$  & 1452.490 & 0.23   \\
 & 20 & $2^+$ & 0.232 & \\
 & 20 & $3^-$ & 0.935 & \\
 \colrule
  & 20 & $0^+$ &  1291.291 & 1.46  \\ 
 
 616 & 20& $1^-$  & 1086.928 &   \\
 & 20 & $2^+$ & 791.117 & \\
 & 20 & $3^-$ & 2.599 & \\
 \colrule
 
  & 20 & $0^+$ &  1288.454 &  \\ 
 
 250 & 20& $1^-$  & 1356.232 & 1.16   \\
 & 20 & $2^+$ & 0.159 & \\
 & 20 & $3^-$ & 52.385 & \\
 
 \colrule
 & 20 & $0^+$ &  1298.718 &  \\ 
 
 350 & 20& $1^-$  & 1351.854 & 1.28  \\
 & 20 & $2^+$ & 8.298 & \\
 & 20 & $3^-$ & 5.147 & \\

 \colrule
  & 20 & $0^+$ &  1296.255 &   \\ 
 
450 & 20& $1^-$  & 1204.854 &  0.08 \\
 & 20 & $2^+$ & 999.783 & \\
 & 20 & $3^-$ & 5.272 & \\

\end{tabular}
\end{ruledtabular}
\end{table}

\begin{figure}[h!]
\begin{center}
\includegraphics[width=0.35\textwidth]{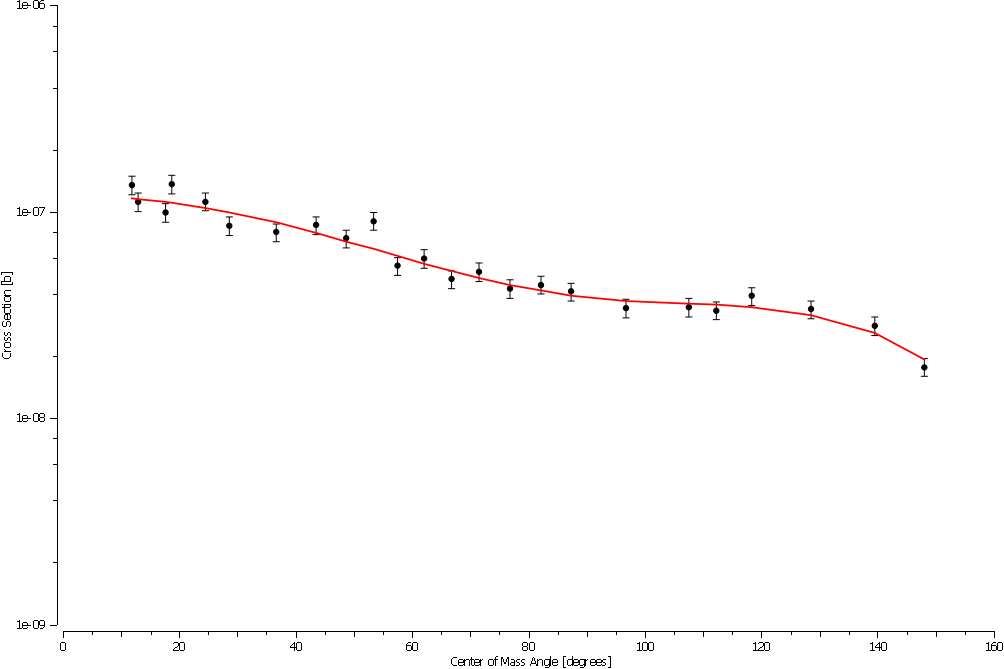}
\includegraphics[width=0.35\textwidth]{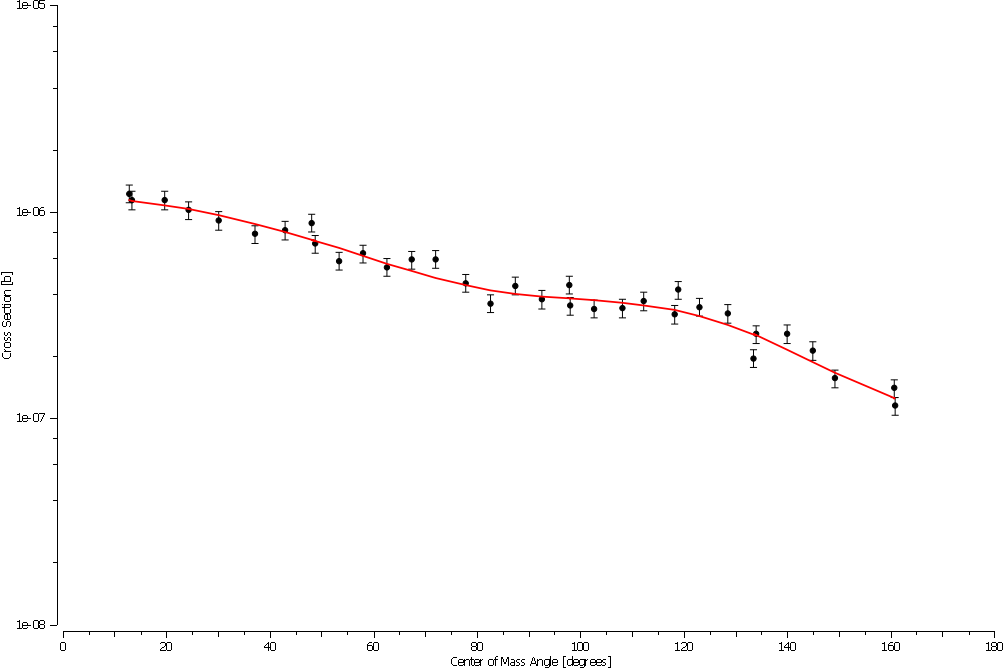}\\
\includegraphics[width=0.35\textwidth]{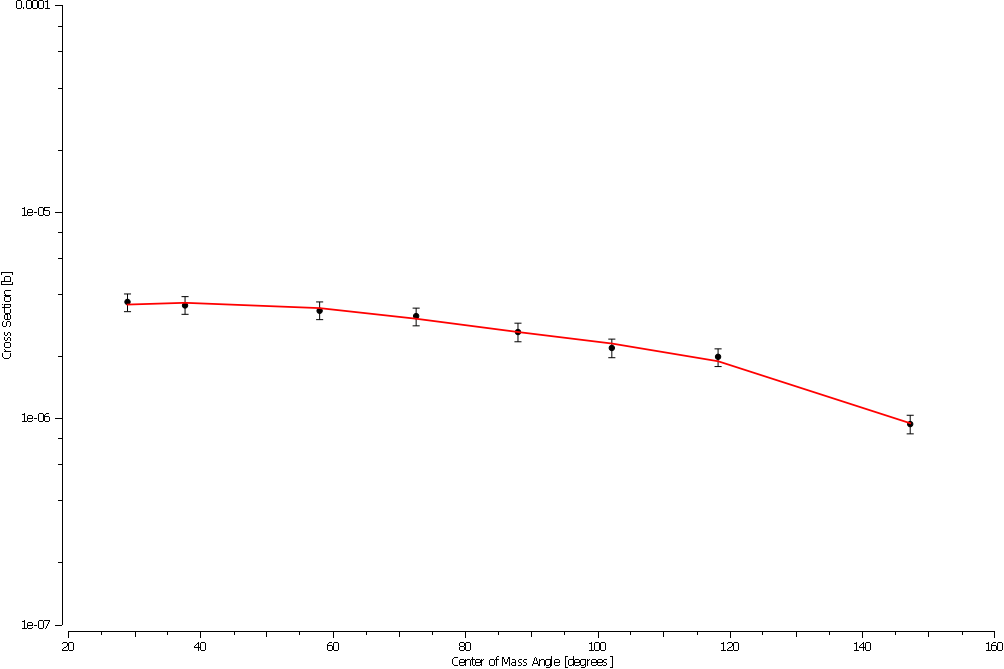}
\caption{R-matrix fitting of Angular distribution for the reaction $^{19}$F(p,$\alpha_0)^{16}O$ at 250, 350 and 450 KeV.The  experimental data are taken from \cite{lw}}.
\label{wirzba et al.}
\end{center}
\end{figure}
\begin{figure}[h!]
\begin{center}
\begin{tabular}{cc}
\includegraphics[width=0.35\textwidth]{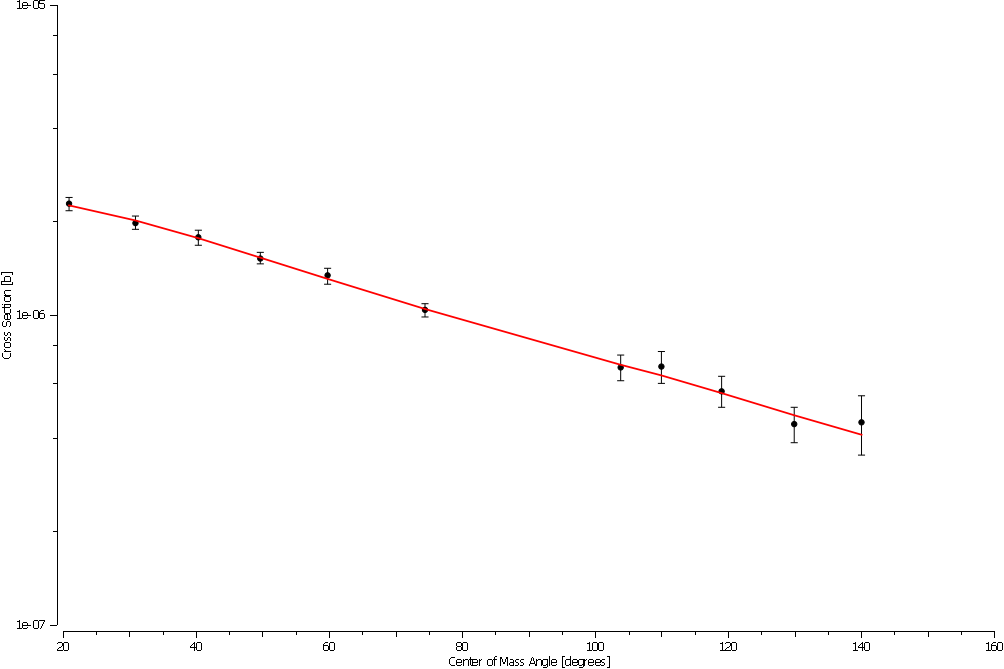}
\includegraphics[width=0.35\textwidth]{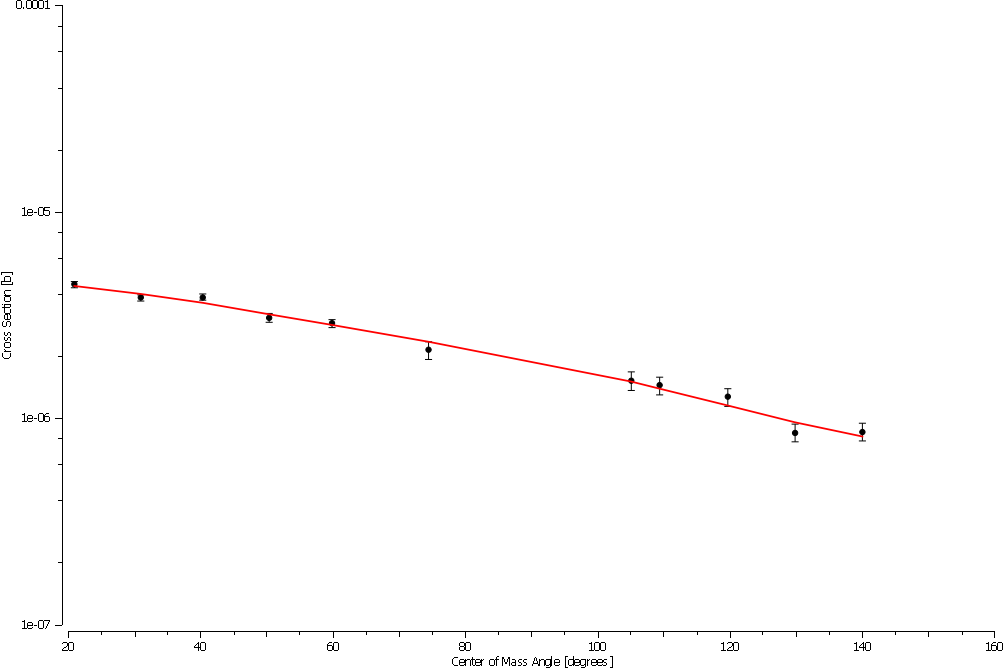}\\
\includegraphics[width=0.35\textwidth]{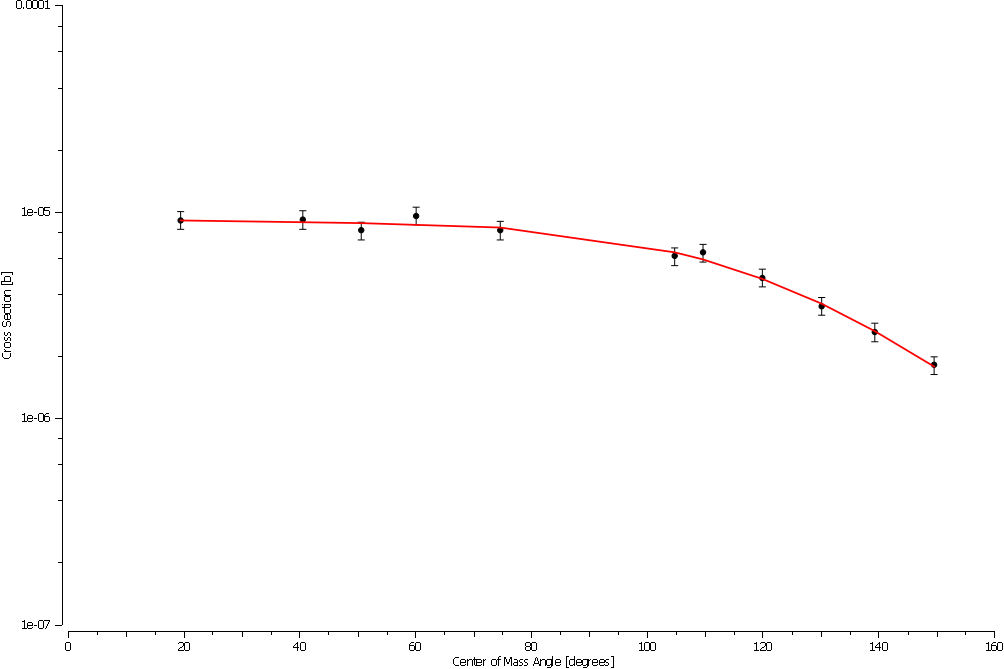}
\includegraphics[width=0.35\textwidth]{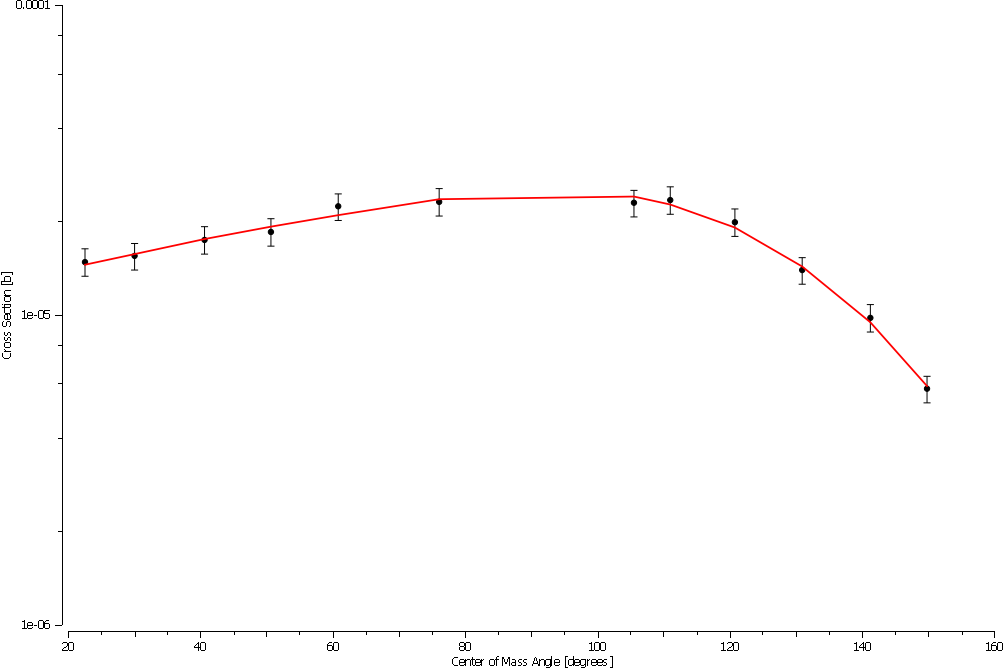}\\
\end{tabular}
\caption{R-matrix fitting of Angular distribution for the reaction $^{19}$F(p,$\alpha_0)^{16}O$ at 327,387,486 and 616 KeV.The  experimental data are taken from \cite{l}}
\label{Lombardo et al.}
\end{center}
\end{figure}

\ \\
\ \\
\ \\
\ \\
\ \\
\ \\
\ \\
\ \\
\ \\
\ \\
\ \\
\ \\
\ \\
\ \\
\ \\
\ \\   

\section{Summary and Conclusions}
In this work, the most recent and lowest energy angular distribution data for the $^{19}$F(p,$\alpha_0$) reaction has been studied in the framework of the DWBA theory. The pickup transfer process explains satisfactorily the data at the lower energies   except the data at 616 keV. The exchange process as suggested by an earlier study is not significant in the alpha emission process. The direct process cannot therefore be neglected in the extraction S-factor as has been done in all previous works. The R-matrix fitting of angular distribution also confirms the contribution of direct part.

\end{document}